\documentclass[a4paper,11pt]{article}
\pdfoutput=1
\usepackage{jcappub}

\usepackage{mathtools}
\usepackage[T1]{fontenc}
\usepackage[normalem]{ulem}

\newcommand{\rR}{\rho_R}
\newcommand{\rp}{\rho_\phi}
\newcommand{\Gp}{\Gamma_\phi}
\newcommand{\gs}{g_\star}
\newcommand{\gss}{g_{\star s}}
\newcommand{\Trh}{T_\text{rh}}
\newcommand{\Tmax}{T_\text{max}}
\newcommand{\Hrh}{H_\text{rh}}
\newcommand{\arh}{a_\text{rh}}
\newcommand{\adm}{a_\text{DM}}
\newcommand{\Br}{\text{Br}}
\newcommand{\mdm}{m_\text{DM}}
\newcommand{\ndm}{n_\text{DM}}
\newcommand{\CC}{\mathcal{C}}

\title{Ultraviolet Freeze-in with a Time-dependent Inflaton Decay}

\author[a]{Basabendu Barman,}
\author[a]{Nicolás Bernal,}
\author[b]{\\Yong Xu,}
\author[c]{and Óscar Zapata}

\affiliation[a]{Centro de Investigaciones, Universidad Antonio Nariño\\Carrera 3 este \# 47A-15, Bogotá, Colombia}
\affiliation[b]{Bethe Center for Theoretical Physics and Physikalisches Institut, Universit\"at Bonn\\ Nussallee~12, 53115 Bonn, Germany}
\affiliation[c]{Instituto de Física, Universidad de Antioquia\\Calle 70 \# 52-21, Apartado Aéreo 1226, Medellín, Colombia}

\emailAdd{basabendu88barman@gmail.com}
\emailAdd{nicolas.bernal@uan.edu.co}
\emailAdd{yongxu@th.physik.uni-bonn.de}
\emailAdd{oalberto.zapata@udea.edu.co}

\abstract{
It is typically assumed that during reheating the inflaton decays with a constant decay width. However, this is not guaranteed and can have a strong impact on the dark matter (DM) genesis.
In the context of the ultraviolet (UV) freeze-in mechanism, if the operators connecting the dark and visible sectors are of sufficiently high mass dimension, the bulk of the DM abundance is produced during and not after reheating.
We study here the impact of a time-dependent decay width of the inflaton on the DM abundance, emphasizing the differences with respect to the cases where the decay is either instantaneous or constant.
We also provide concrete examples for DM production via UV freeze-in, e.g., from 2-to-2 scatterings of standard model particles, or from inflaton scatterings or decays, elucidating how the time-dependence influences the DM yield.
\vspace{2cm}
\begin{center}
{\it ``All war is a symptom of man's failure as a thinking animal.''\\-John Steinbeck, `Once There Was A War' (1958)}
\end{center}
}

\begin{document}
\begin{flushright}
    PI/UAN-2022-710FT
\end{flushright}

\maketitle

\section{Introduction}
The known baryonic matter can only explain $\sim 15\%$ of the total matter budget of the Universe~\cite{Planck:2018vyg}; the rest is referred as dark matter (DM)~\cite{Bertone:2016nfn, deSwart:2017heh}, and is currently one of the most profound mysteries in both particle physics and cosmology. Regarding the nature of DM, the weakly interacting massive particles (WIMPs) are one of the most prominent candidates~\cite{Steigman:1984ac} (for a review, see e.g. Refs.~\cite{Jungman:1995df, Bertone:2004pz, Feng:2010gw}). In the WIMP scenario, DM particles carry an interaction strength at the electroweak scale, which allow them to thermalize with the baryon-photon plasma in the early Universe and then freeze out, reproducing the observed relic density $\Omega_{\rm DM} h^2 \simeq 0.12$~\cite{Planck:2018vyg}. This scenario is very appealing since several extensions of the standard model (SM) of particle physics naturally include WIMPs. However, currently the strong observational constraints on the typical WIMP parameter space motivate quests beyond this paradigm (see, e.g., Refs.~\cite{Roszkowski:2017nbc, Arcadi:2017kky}).

Alternatively to WIMPs, feebly interacting massive particles (FIMPs), which couple to the SM sector very feebly evading therefore the current experimental constraints, have been attracting heated discussion recently~\cite{McDonald:2001vt, Choi:2005vq, Kusenko:2006rh, Petraki:2007gq, Hall:2009bx, Bernal:2017kxu}. In the early Universe, FIMPs can be generated from either the decay or annihilation of states in the visible sector. When the SM temperature becomes smaller than the typical mass scale of the interaction (i.e. the maximum of the DM and the mediator mass), the generation process becomes Boltzmann suppressed, giving rise to a constant comoving DM number density; such a scenario is referred as the freeze-in mechanism~\cite{Hall:2009bx}.

The FIMP paradigm requires very suppressed interaction rates between the dark and visible sectors, which can be achieved either in its infrared version via small couplings (typically of the order $\sim 10^{-11}$), or in its ultraviolet (UV) version via non-renormalizable operators~\cite{Elahi:2014fsa}, suppressed by a high mass scale.
The latter scenario is particularly interesting, as the DM yield is sensitive to the highest temperature $\Tmax$ reached by the SM plasma~\cite{Giudice:2000ex}, controlled by the dynamics of the inflaton decay.
In the sudden decay approximation for the inflaton, $\Tmax$ corresponds to the reheating temperature $\Trh$, characterizing the onset of the radiation-dominance era.
However, once away from the instantaneous reheating, $\Tmax$ can be significantly larger than $\Trh$.

It is interesting to note that both $\Tmax$ and $\Trh$ are controlled by the inflaton dynamics during reheating, in particular by its dissipation rate $\Gp$, typically assumed to be constant. 
However, this should not be the case.
In scenarios where the inflaton decays via higher-order operators or oscillates around a potential steeper than quadratic, its decay width may not be constant, but features a time dependence~\cite{Bodeker:2006ij, Mukaida:2012qn,  Co:2020xaf, Garcia:2020wiy, Ahmed:2021fvt, Banerjee:2022fiw}. 
In particular, it was recently shown that in contrast to the conventional case where during reheating the SM temperature scales as $T(a) \propto a^{-3/8}$ (where $a$ corresponds to the scale factor), the SM temperature could show a non-trivial dependence with the scale factor~\cite{Co:2020xaf}.

Without loss of generality, we parametrize the dissipation rate as $\Gp(a,\, T) \propto a^k\, T^q$. Note that the conventional result with constant $\Gp$ corresponds to the case where $q = k = 0$. In general, one may expect a varying decay rate.
For example, the dynamics of a coherently oscillating scalar field in the early Universe can be affected by the thermal environment due to thermal modification to the effective potential, non-perturbative particle production or non-topological effects~\cite{Mukaida:2012qn}. If $\phi$ oscillates with a zero-temperature mass effective potential $V \propto m_\phi^2\, \phi^2$, it is possible to have $\Gp\propto T$, in the limit the effective (temperature corrected) mass $m_\phi^\text{eff}\sim m_\phi\ll T$~\cite{Mukaida:2012qn}. In these cases, it is possible to realize $k = 0$ and $q = 1$. 
On the other hand, if the inflaton oscillates in the vicinity of a potential steeper than quadratic, the decay rate also typically features a scale-factor dependence due to the field dependence of inflaton mass~\cite{Garcia:2020wiy}. This can lead to a variety of values of $k$  depending on the spin of the decay products%
\footnote{Note that $\Gp \propto m_\phi$ or $1/m_\phi$ for fermions or bosons in the final state, respectively.} and the shape of the inflaton potential during reheating. For example, for reheating in a quartic potential, one has $\Gp \propto a^{3/4}$ $(a^{-3/4})$ if the inflaton decays to bosons (fermions)~\cite{Garcia:2020wiy}; this corresponds to $q=0$ and $k=3/4$, or $q=0$ and $k=-3/4$. We consider a general parametrization for the decay rate which captures a variety of dynamics during reheating, including the effect of $i)$ decays via higher-order operators ~\cite{Co:2020xaf, Moroi:2020has}, $ii)$ shapes of inflaton potentials during reheating~\cite{Garcia:2020wiy, Daido:2017wwb}, as well as $iii)$ feedback from the thermal background~\cite{Mukaida:2012qn, Mukaida:2015nos}.

The nontrivial behavior of the SM temperature during reheating leads to a diverse DM phenomenology, in particular for those scenarios where its yield is sensitive to the (highest) cosmic temperature after inflation.
In particular, if the operators connecting the dark and visible sectors are of sufficiently high mass dimension, the bulk of the DM could be produced during and not after reheating. This has been characterized by defining a ``boost factor'' $B$ for the DM relic density, which is the ratio of the DM abundance taking into account non-instantaneous reheating relative to the abundance in the sudden decay approximation~\cite{Garcia:2017tuj}.%
\footnote{Another boost factor could be defined due to the effects of thermalization and number-changing processes in the dark sector.
They can also have a strong impact, in particular enhancing the DM relic abundance by several orders of magnitude~\cite{Bernal:2020gzm}. For the sake of simplicity we are assuming here no sizable self-interactions within the dark sector.}
It was initially pointed out that such boosts only depend on the mass dimension of the operator and the ratio $\Tmax/\Trh$, however the equation-of-state parameter of the inflaton during reheating also plays a major role~\cite{Bernal:2019mhf, Garcia:2020eof, Bernal:2020bfj, Allahverdi:2020bys}.
Subsequent papers have explored the impact of this boost factor in specific models~\cite{Chen:2017kvz, Bernal:2018qlk, Bhattacharyya:2018evo, Chowdhury:2018tzw, Kaneta:2019zgw, Banerjee:2019asa, Chanda:2019xyl, Dutra:2019xet, Dutra:2019nhh, Mahanta:2019sfo, Cosme:2020mck, Bernal:2020fvw, Bernal:2020qyu, Bernal:2020yqg, Bernal:2021qrl}.

In this work we demonstrate that the phenomenology of the UV freeze-in paradigm strongly depends on the dynamics of the inflaton decay during reheating. Using a general parametrization for the inflaton decay width, we carefully study the DM production during reheating and compute the corresponding boost factors, by comparing it to the approximation  where the inflaton decays suddenly. In a completely model-independent fashion, we show that the DM yield can have a power-law boost in the ratio $\Tmax/\Trh$, particularly important if the SM temperature drops fast enough during reheating. We typically concentrate on the situation where the initial radiation density is negligibly small, and the total energy density of the Universe is dominated by the inflaton during reheating.
In such a framework, we find that the nontrivial decay dynamics of the inflaton, and consequently the boost factor, is completely determined by some combination of the exponents $q$ and $k$, together with the two Hubble scales $H_I$ and $\Hrh$ defined at the beginning and the end of the reheating epoch,  respectively. Finally, we provide some realistic examples of DM production in the early Universe, namely, gravitational UV freeze-in, inflaton scattering, and inflaton decay, to demonstrate how a time-dependent inflaton decay width can influence the DM yield in each of these cases.

The paper is organized as follows. In Sect.~\ref{sec:sudden-decay}, we briefly revisit the phenomenology of UV freeze-in where DM is produced out of annihilations of SM particles, in the conventional approximation in which the inflaton suddenly decays.  In Sect.~\ref{sec:non-inst-decay}, we solve the Boltzmann equation to compute the background evolution in the presence of a time-dependent inflaton decay rate. The phenomenology of the modified FIMP DM yield is studied in Sect.~\ref{sec:uvFreeze-in}.  In Sec.~\ref{sec:examples} we provide three concrete instances to exemplify our results. Finally, we summarize our findings in Sect.~\ref{sec:concl}.

\section{UV Freeze-in in the Sudden Decay Approximation}
\label{sec:sudden-decay}
The evolution of the DM number density $\ndm$ is governed by a Boltzmann equation (BEQ), that can be written in a generalized form as
\begin{equation} \label{eq:BE0}
    \frac{d\ndm}{dt} + 3\, H\, \ndm = \gamma(T)\,,
\end{equation}
where $\gamma(T)$ corresponds to the DM production rate density out of SM particles, as a function of the bath temperature $T$.
In case where the Universe energy density is dominated by SM radiation, the Hubble expansion rate $H$ is given by
\begin{equation}\label{eq:hub-rad}
    H(T) = \sqrt{\frac{\rR(T)}{3\, M_P^2}}\,,
\end{equation}
with the SM radiation energy density
\begin{equation}\label{eq:rho-rad}
    \rR(T) = \frac{\pi^2}{30}\, \gs(T)\, T^4\,,
\end{equation}
where $g_\star$ is the number of relativistic degrees of freedom contributing to the SM energy density~\cite{Drees:2015exa}, and $M_P$ the reduced Planck mass.
In the case where the SM entropy is conserved, it is instructive to solve Eq.~\eqref{eq:BE0} in terms of the DM yield $Y\equiv \ndm /s$, defined as a ratio of DM number to entropy density of the Universe, where $s(T) \equiv \frac{2\pi^2}{45}\gss\, T^3$, with $\gss(T)$ being the number of relativistic degrees of freedom contributing to the SM entropy~\cite{Drees:2015exa}.
At temperatures much higher than the electroweak scale, $\gs = \gss = 106.75$.
On substituting in Eq.~\eqref{eq:BE0} one obtains
\begin{equation}\label{eq:BEQ-Y}
    \frac{dY}{dT} = -\frac{\gamma(T)}{H(T)\, T\, s(T)}\,. 
\end{equation}

Now, in case of UV freeze-in, the temperature $T$ of the thermal bath is always lower than the typical mass scale of the interaction, implying that the operators through which the DM communicates with the visible sector are non-renormalizable. In such a scenario, the reaction rate for DM produced from the SM bath can be parametrized as~\cite{Elahi:2014fsa}
\begin{equation}\label{eq:gamma-SM}
    \gamma(T) = \frac{T^n}{\Lambda^{n-4}}\,,
\end{equation}
where $\Lambda$ is a dimensionful parameter which is the effective interaction scale between the DM and the SM, coming from an operator of mass dimension $d = 2 + n/2$, (with $n \geq 6$).\footnote{Detection prospects of UV freeze-in via non-renormalizable operators in an early matter dominated era have been discussed in Ref.~\cite{Calibbi:2021fld}.}
It is thus implied that the effective description is valid at temperatures $T < \Lambda$.

In the approximation where the inflaton decays instantaneously, the reheating temperature $\Trh$ corresponds to the maximum temperature achieved by the thermal bath and characterizes the onset of the radiation-dominated era.
For SM temperatures much smaller than $\Trh$ (i.e., $T \ll \Trh$), it is possible to solve Eq.~\eqref{eq:BEQ-Y} analytically, obtaining
\begin{equation} \label{eq:Y0}
    Y_0 \equiv Y(T\ll \Trh) = \frac{135}{2\pi^3 (n-5)\, \gss} \sqrt{\frac{10}{\gs}}\, \frac{M_P\, \Trh^{n-5}}{\Lambda^{n-4}}\,,
\end{equation}
where $Y_0$ denotes the DM yield at the end of reheating and assuming $\mdm \ll \Trh$, $\mdm$ being the DM mass. It is important to emphasize that we have also assumed a DM abundance initially negligible and a production cross section sufficiently small such that DM remains out of chemical equilibrium with the SM bath. Finally, to match the whole observed abundance, the DM yield has to be fixed so that $\mdm\, Y_0 = \Omega_\text{DM} h^2 \frac{1}{s_0}\, \frac{\rho_c}{h^2} \simeq 4.3 \times 10^{-10}$~GeV, where $\mdm$ is the DM mass in GeV, $\rho_c \simeq 1.1 \times 10^{-5} h^2$~GeV/cm$^3$ is the critical energy density, $s_0 \simeq 2.9 \times 10^3$~cm$^{-3}$ is the entropy density at present, and $\Omega_\text{DM} h^2 \simeq 0.12$~\cite{Planck:2018vyg}.

\section{Non-instantaneous Reheating with a Time-dependent Inflaton Decay}
\label{sec:non-inst-decay}
Although reheating is commonly approximated to be sudden, in reality, the decay of the inflaton is non-instantaneous and typically characterized by an exponential decay law~\cite{Scherrer:1984fd}.
Therefore, the evolution of inflaton and radiation energy densities ($\rp$ and $\rR$, respectively) is governed by a set of coupled BEQs that read~\cite{Giudice:2000ex}%
\footnote{We would like to mention that  we have focused on a scenario where inflaton decays perturbatively. Potentially relevant non-perturbative preheating processes~\cite{Kofman:1997yn} will not be studied here.}
\begin{align}
    &\frac{d\rp}{dt} + 3\, H\, \rp = -\Gp\,\rp\,,\label{eq:cBEQ1}\\
    &\frac{d\rR}{dt} + 4\, H\, \rR = +\Gp\, \rp\,,\label{eq:cBEQ2}
\end{align}
with, in this case, the Hubble expansion rate $H$ is given by
\begin{equation}\label{eq:tot-hub}
    H^2 = \frac{\rR + \rp}{3\, M_P^2}\,,
\end{equation}
and $\Gp$ being the total perturbative decay width of the inflaton. It is also assumed that during reheating (i.e., for $a_I< a < \arh$, where $a_I$ and $\arh$ correspond to the scale factors at the end of inflation and at the end of reheating, respectively), the inflaton energy density scales as non-relativistic matter, i.e., $\rp(a) \propto a^{-3}$. This scaling is characteristic of an inflaton oscillating in a quadratic potential.
Therefore, during reheating
\begin{equation}\label{eq:rho-inf}
    \rp(a) \simeq 3 M_P^2\, H_I^2 \left(\frac{a_I}{a}\right)^3,
\end{equation}
with $H_I\equiv H(a = a_I)$ denoting the Hubble parameter at the end of inflation. As advocated earlier, generalizing the widely used assumption of constant $\Gp$, here we parametrize the inflaton decay as a function of temperature and scale factor as
\begin{equation}\label{eq:inf-decay}
    \Gp(a,\, T) = \CC \left(\frac{a}{\arh}\right)^k \left(\frac{T}{\Trh}\right)^q \Hrh\,,
\end{equation}
with $\CC = \CC(k,\, q)$ a constant that will be conveniently fixed in the following.%
\footnote{Note that this parametrization slightly differs from the one in Ref.~\cite{Co:2020xaf}.} 
We would like to stress that the previous equation is just a convenient parametrization, the exact expression could be computed once a full model is fixed.
For instance, the present parametrization is not valid before the thermalization of the SM bath.
The decay width can be recasted in terms of the radiation energy density following Eq.~\eqref{eq:rho-rad} as
\begin{equation}
    \Gp(a) = \CC \left(\frac{30}{\pi^2\, \gs}\right)^{q/4}\left(\frac{a}{\arh}\right)^k \left(\frac{\rR^{1/4}(a)}{\Trh}\right)^q \Hrh\,,
\end{equation}
where $q$ and $k$ are parameters given by the specific coupling of the inflaton to SM particles.
Note that if $k = q = 0$, one actually reproduces the instantaneous reheating scenario with $\Gp(\Trh)=\CC\, \Hrh$. In that sense, the parameterization of $\Gp$ in Eq.~\eqref{eq:inf-decay} is generic and captures both the standard and some potentially unusual features of the reheating dynamics.
The reheating period ends at the so-called reheat temperature $\Trh$, which can be defined using $\rR(\Trh) \equiv 3 M_P^2\, \Hrh^2$,
\begin{equation} \label{eq:Trh}
    \Trh^2 = \frac{3}{\pi}\sqrt{\frac{10}{\gs}}\, M_P\, \Hrh\,,
\end{equation}
which also marks the onset of the radiation-domination era.

During reheating, it is possible to analytically solve Eqs.~\eqref{eq:cBEQ1} and~\eqref{eq:cBEQ2} in order to extract the radiation energy density as
\begin{align}\label{eq:rhoR}
     \rho_R(a) &= \Biggl[\left(3\, M_P^2\, \Hrh^2\right)^\frac{4-q}{4} \frac{\CC}{2} \frac{4-q}{5+2k-2q}\, \frac{H_I}{\Hrh}  \left(\frac{a_I}{\arh}\right)^k \left(\Bigl(\frac{a}{a_I}\Bigr)^\frac{5+2k-2q}{2}-1\right)\nonumber\\
     &\qquad + \rho_R(a_I)^{\frac{4-q}{4}}\Biggr]^\frac{4}{4-q}\,\left(\frac{a_I}{a}\right)^4,
\end{align}
where it was assumed that the total energy density is dominated by inflatons during reheating.
As just after inflation one expects a completely subdominant SM radiation energy density, that is $\rR(a_I) \simeq 0$, here we will focus on the case $q < 4$. For completeness, however, we would like to mention the consequences of having $q>4$. As discussed in Ref.~\cite{Co:2020xaf}, although the $q>4$ scenario is governed by the same Eq.~\eqref{eq:rhoR}, it gives rise to a very different outcome. This is typically due to the fact that unlike the $q<4$ scenario, the initial radiation energy density is no more negligible here. Now, for $5-2q+2k>0$, the temperature continues to decrease during reheating, with a sharp increase towards the end of reheating showing a sudden completion of dissipation. On the other hand, with $5-2q+2k<0$, the dissipation is never complete and one ends up with the usual instantaneous reheating. Now, on top of $q<4$, we additionally require $5-2q+2k > 0$ to guarantee an efficient energy transfer from the inflaton to the SM radiation.
Under such considerations one has
\begin{equation} \label{eq:rhoR2}
    \rR(a) \simeq 3\, M_P^2\, \Hrh^2 \left[\frac{\CC}{2}\frac{4-q}{5 - 2q + 2k}\, \frac{H_I}{\Hrh} \left(\frac{a_I}{\arh}\right)^k\right]^\frac{4}{4-q} \left(\frac{a_I}{a}\right)^\frac{6-4k}{4-q}.
\end{equation}
At the end of reheating (i.e., at $a = \arh$), the inflaton and the radiation energy densities are equal, $\rR(\arh) = \rp(\arh) = 3 M_P^2\, \Hrh^2$, and therefore Eq.~\eqref{eq:rhoR2} implies
\begin{equation}\label{eq:aIarh}
    \frac{a_I}{\arh} = \left(\frac{2}{\CC}\, \frac{5+2k-2q}{(4-q)}\, \frac{\Hrh}{H_I}\right)^\frac23 \equiv \left(\frac{\Hrh}{H_I}\right)^\frac23,
\end{equation}
where in the last term we have used the Hubble-scale factor relation during reheating. The constant $\CC$ is thus fixed from now on as
\begin{equation}\label{eq:CC}
    \CC = \frac{2\,\left(5 - 2q + 2k\right)}{4-q}\,,
\end{equation}
which in turn allows to rewrite Eq.~\eqref{eq:rhoR2} as
\begin{equation}\label{eq:rhoR-approx}
    \rR(a) \simeq 3\, M_P^2\, \Hrh^2 \left(\frac{\arh}{a}\right)^\frac{2\, (3-2k)}{4-q}.
\end{equation}
It is worth mentioning that for $k > 3/2$, $\rR$ and therefore the SM temperature continuously increases during reheating, and therefore $\Trh$ corresponds to the maximum temperature reached by the SM plasma.
Additionally, if $k=3/2$, the SM temperature remains constant during the inflaton-dominated era. Finally, note that for $k < 3/2$, in the first stages of reheating, the SM bath reaches a maximal temperature of $\Tmax$~\cite{Giudice:2000ex}, given by
\begin{equation} \label{eq:Tmax}
    \Tmax = \Trh  \left(\frac{H_I}{\Hrh}\right)^\frac{3-2k}{3\, (4-q)},
\end{equation}
which is higher than $\Trh$. 

\begin{figure}[t!]
    \def\sepf{0.39}
	\centering
    \includegraphics[scale=\sepf]{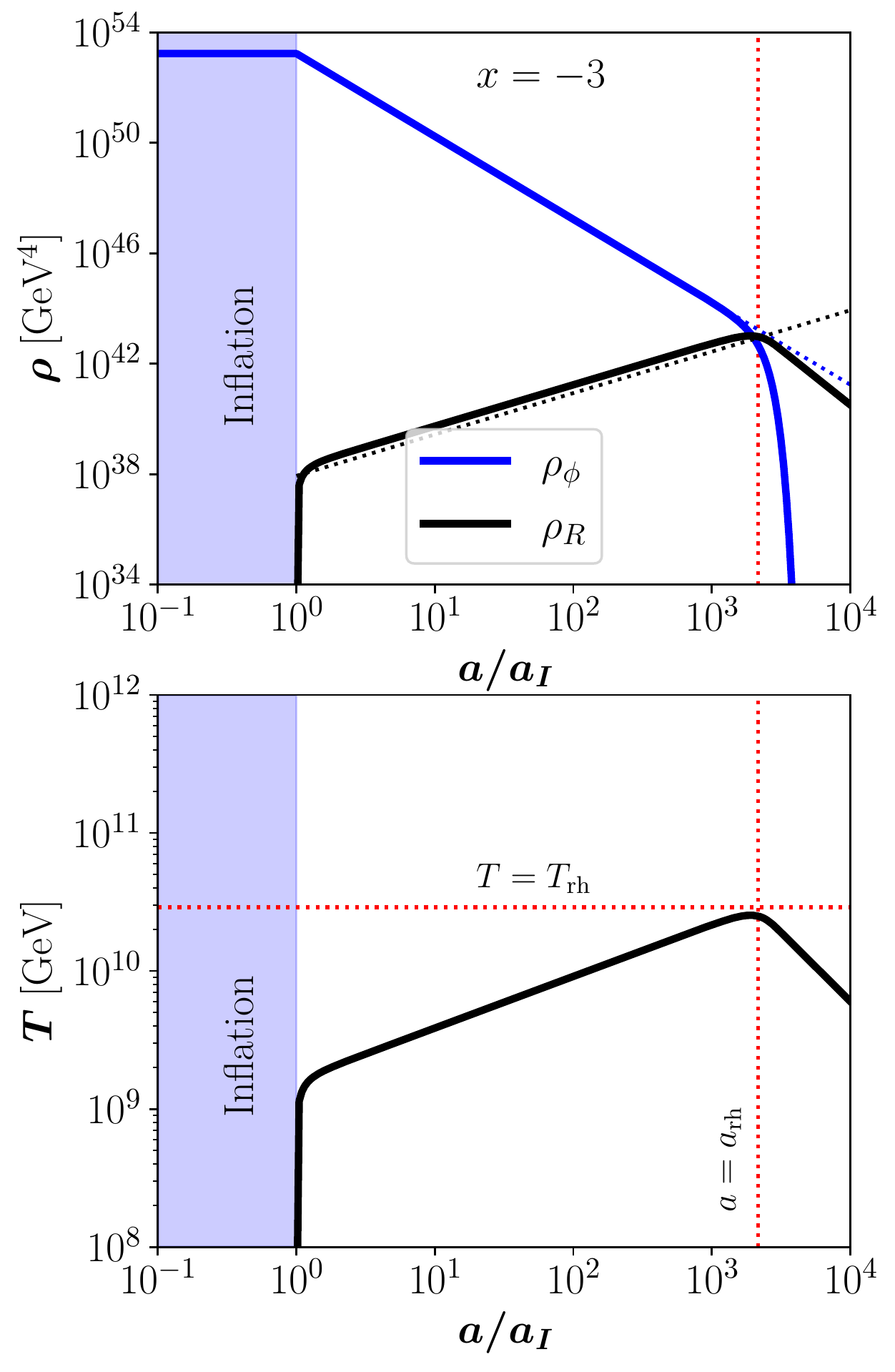}
    \includegraphics[scale=\sepf]{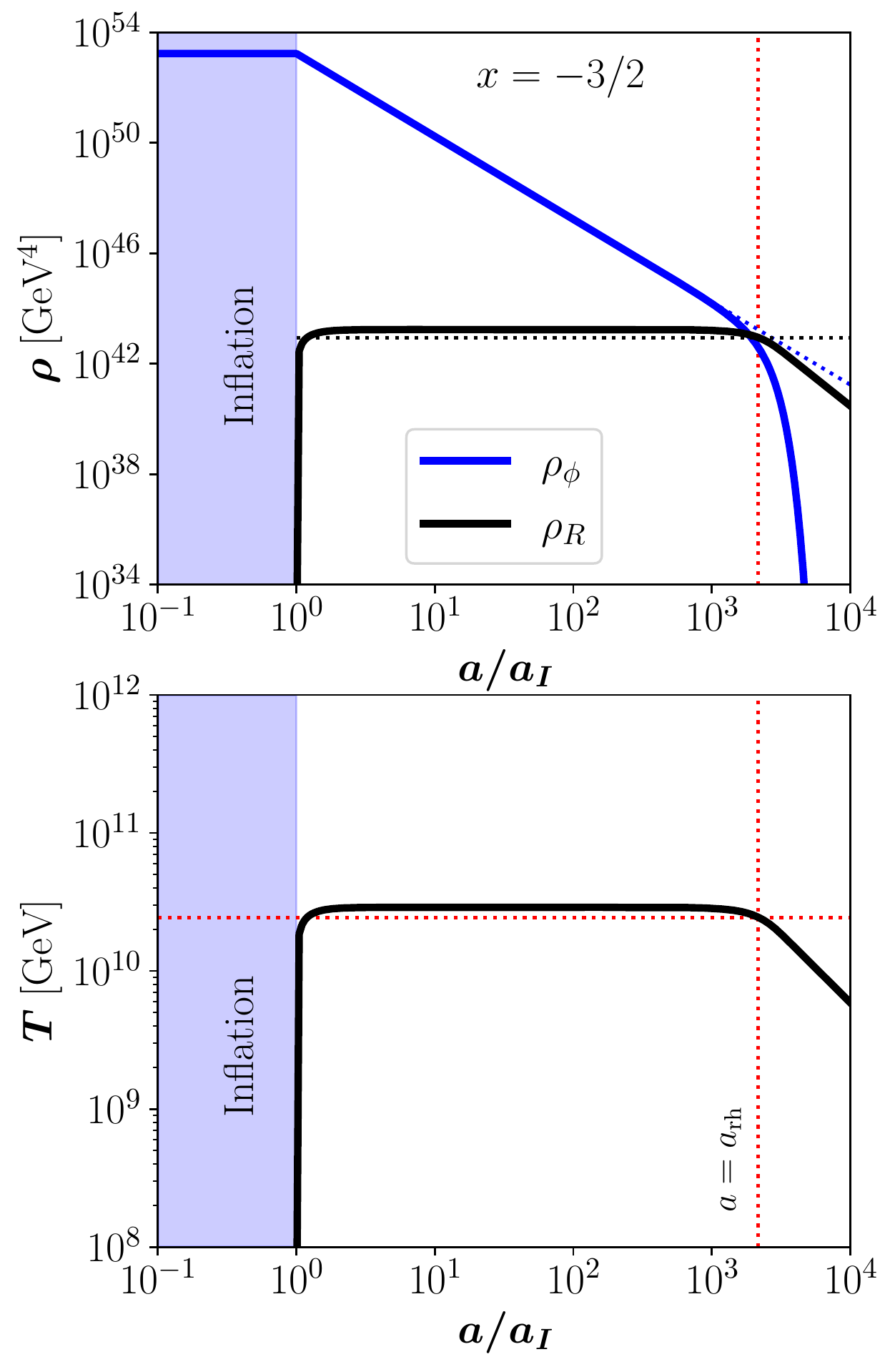}
    \includegraphics[scale=\sepf]{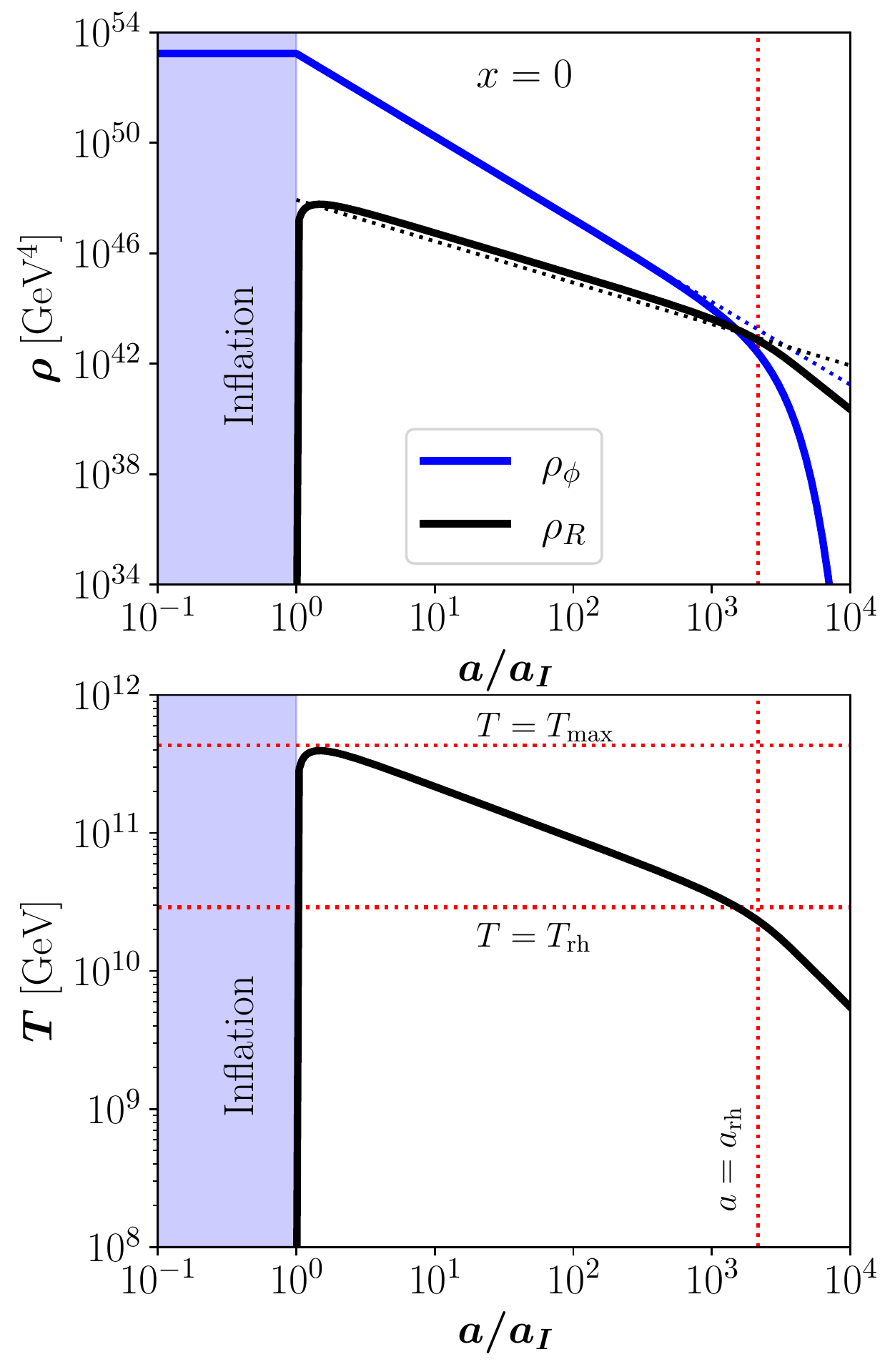}
    \caption{Upper panels: Evolution of the inflaton (blue) and SM radiation (black) energy densities as a function of the scale factor, for $x=-3$ (left), $x=-3/2$ (center) and $x=0$ (right), taking $H_I = 10^8$~GeV and $\Hrh = 10^3$~GeV.
    Lower panels: The corresponding evolution of the photon temperature.
    }
	\label{fig:rho}
\end{figure}

It is interesting to note that the decay width could be simplified to the form
\begin{equation}
    \Gp(a) \simeq \frac{5 - 2x}{2}\, \Hrh \left(\frac{\arh}{a}\right)^x,
\end{equation}
where
\begin{equation}
    x \equiv \frac{3q-8k}{2\, (4-q)}\,,
\end{equation}
with $x < 5/2$ coming from the condition $5-2q+2k > 0$. Additionally, one still has to guarantee that $q < 4$.
We emphasize that a constant decay width is reached in the case $k=q=0$, but also more generally for $x=0$, corresponding to $k = \frac38\, q$.
Cases $x<0$ and $x>0$ generate an inflaton decay width that increases or decreases with time, respectively. 
The SM radiation energy density becomes, in turn,
\begin{equation}\label{eq:rhoR3}
    \rR(a) \simeq 3\, M_P^2\, \Hrh^2 \left(\frac{\arh}{a}\right)^\frac{3+2x}{2}.
\end{equation}
Before moving on, we would like to emphasize that, as expected, the present scenario demands a set of {\it three} free parameters that could conveniently be chosen to be $x$, together with the two Hubble scales: $H_I$ and $\Hrh$. The Planck collaboration puts an upper limit on the inflationary scale $H_I \leq 2.5 \times 10^{-5}~M_P$~\cite{Akrami:2018odb} from the non-observation of CMB tensor modes. On the other hand, by combining light element abundance measurements with CMB and large-scale structure data, a fairly robust lower limit on the reheating temperature $\Trh \gtrsim 4$~MeV can be set at 95\%~CL~\cite{Hannestad:2004px, deSalas:2015glj, Hasegawa:2019jsa} that in turn puts a lower bound on $\Hrh \gtrsim 3.0 \times 10^{-42}~M_P$.
In the upper panels of Fig.~\ref{fig:rho} the evolution of the inflaton (blue) and SM radiation (black) energy densities as a function of the scale factor, for $x = -3$ (left), $x = -3/2$ (center) and $x = 0$ (right).
We take $H_I = 10^8$~GeV and $\Hrh = 10^3$~GeV.
The lower panels show the corresponding evolution of the photon temperature. The thick solid lines show the numerical solutions of the system of Boltzmann equations, whereas the red dotted lines the analytical approximations.

\section{UV Freeze-in with a Time-dependent Inflaton Decay}
\label{sec:uvFreeze-in}
In this section, we address the effect of a generic time-dependent inflaton decay rate on the DM production via UV freeze-in.%
\footnote{We emphasize that here we are focusing on a non-instantaneous decay of the inflaton, but on an instantaneous thermalization of the SM plasma~\cite{Harigaya:2014waa, Mukaida:2015ria, Harigaya:2019tzu}.
The SM particles do not necessarily thermalize suddenly, and thus they could be initially distributed with smaller occupation numbers and harder momenta~\cite{Ellis:2015jpg, Harigaya:2013vwa, Garcia:2018wtq, Drees:2021lbm}.}
Since the SM entropy is not conserved during reheating, it is more convenient to rewrite  Eq.~\eqref{eq:BE0} in term of the comoving DM number density $N \equiv \ndm \times a^3$ as%
\footnote{We note that DM could also be produced via freeze-in from parametric resonances (i.e. preheating)~\cite{Garcia:2021iag}.}
\begin{equation} \label{eq:BEa}
    \frac{dN}{da} = \frac{a^2 \gamma}{H} \simeq \left(\frac{\arh^{3\, n/n_c}}{a_I}\right)^{3/2} \frac{\Trh^n}{H_I\, \Lambda^{n-4}}\, a^{\frac72 - \frac92 \frac{n}{n_c}},
\end{equation}
with
\begin{equation} \label{eq:nc}
    n_c \equiv \frac{36}{3 + 2x} = 9\, \frac{4-q}{3-2k}\,,
\end{equation}
defined for $x \neq -3/2$ (or equivalently, for $k \ne 3/2$). The case $x = -3/2$, corresponding to a constant SM temperature during reheating, will be analyzed later on.

Now, the DM number at the end of reheating can be analytically obtained by integrating Eq.~\eqref{eq:BEa} over the range $a_I < a < \arh$,
\begin{equation}
    N(\arh) \simeq 
        \begin{dcases}
            \frac29 \frac{n_c}{n_c - n}\, \frac{\arh^3\,  \Trh^n}{H_I\, \Lambda^{n-4}} \left(\frac{\arh}{a_I}\right)^\frac32 \left[1 - \left(\frac{a_I}{\arh}\right)^{\frac92 \frac{n_c-n}{n_c}}\right] & \text{ for } n\neq n_c\,,\\
            \frac{\arh^3\, \Trh^{n_c}}{H_I\, \Lambda^{n_c-4}}\, \left(\frac{\arh}{a_I}\right)^\frac32 \ln\frac{\arh}{a_I} & \text{ for } n=n_c\,,
        \end{dcases}
\end{equation}
where a DM lighter than $\Trh$ was assumed.
The case where DM is heavier than $\Trh$, but still lighter than $\Tmax$, will be considered separately. It is important to note that we are also taking a subdominant initial DM number density, $N(a_I)\simeq 0$, a common assumption at the end of inflation.
Although the SM entropy density is not conserved when the inflaton is decaying, one can further define the DM yield at $a=\arh$, namely $Y(\arh) = \ndm(\arh)/s(\Trh) = N(\arh)/(\arh^3\, s(\Trh))$ as
\begin{equation} \label{eq:Yrh}
    Y(\arh) \simeq 
        \begin{dcases}
            \frac{n_c}{n-n_c}\, \frac{5}{\pi^2\, \gss}\, \frac{\Trh^{n-3}}{\Hrh\, \Lambda^{n-4}}\, \Biggl[\left(\frac{H_I}{\Hrh}\right)^{3\,\frac{n-n_c}{n_c}} - 1\Biggr] & \text{ for } n\neq n_c\,,\\
            \frac{45}{(2\, \pi^2)\, \gss}\, \frac{\Trh^{n_c}}{\Trh^3\, \Hrh\, \Lambda^{n_c-4}}\, \ln\left(\frac{H_I}{\Hrh}\right)^\frac23\, & \text{ for } n=n_c\,.
        \end{dcases}
\end{equation}
After the end of reheating, the SM entropy is conserved and therefore $Y(\arh)$ remains constant.

To quantify the DM production during reheating, we define a boost factor $B$ for the DM relic density, which is the ratio of the DM abundance taking into account non-instantaneous reheating relative to the abundance under the sudden decay approximation~\cite{Garcia:2017tuj, Bernal:2019mhf}. Therefore, the total DM yield $Y_\text{total}$ corresponds to the production after and during reheating, so that
\begin{equation}
    Y_\text{total} = Y_0 + Y(\arh) = Y_0 \times \left(1 + B\right)\,.
\end{equation}
By comparing Eqs.~\eqref{eq:Y0} with~\eqref{eq:Yrh}, the boost factor is estimated to be
\begin{equation}\label{eq:boost}
    B \equiv \frac{Y(\arh)}{Y_0} \simeq 
        \begin{dcases}
            \frac29\, \frac{(n-5)\,n_c}{n-n_c} \left(\frac{H_I}{\Hrh}\right)^{3\,\frac{n-n_c}{n_c}} & \text{ for } n > n_c\,,\\
            \frac23\, (n_c-5)\,\ln\left(\frac{H_I}{\Hrh}\right)\, & \text{ for } n=n_c\,,\\
            \frac29\, \frac{(n-5)\,n_c}{n_c-n} & \text{ for } n < n_c\,,
        \end{dcases}
\end{equation}
which tells that a large boost (i.e., a power-law enhancement in the ratio $H_I/\Hrh$) in the DM production during reheating appears for the case where $n > n_c$. If $n=n_c$, the boost is logarithmic in the ratio $H_I/\Hrh$, whereas for $n< n_c$ an $\mathcal{O}(1)$ boost is expected. We note that such a boost depends on $x$, $n$, and the ratio of the Hubble expansion rates, but not on $\Lambda$ or $\mdm$. The case $x=0$ (i.e., an inflaton with an energy density that during reheating scales like non-relativistic matter and decays with a constant decay width) matches with previous results reported in the literature~\cite{Garcia:2017tuj, Bernal:2019mhf}.
Note that in the case with $x > -3/2$ (or equivalently  $k < 3/2$), the SM energy density monotonically decreases during reheating, so  there is a well defined $\Tmax$, given by Eq.~\eqref{eq:Tmax}. In this case, one can further write the boost factor in Eq.~\eqref{eq:boost} in terms of $\Tmax/\Trh$ as
\begin{equation} \label{eq:boost2}
  B \simeq 
        \begin{dcases}
            \frac29\, \frac{(n-5)\,n_c}{n-n_c} \left(\frac{\Tmax}{\Trh} \right)^{n-n_c} & \text{ for } n > n_c\,,\\
            \frac29\,n_c\, (n_c-5)\,\ln\left(\frac{\Tmax}{\Trh} \right)\, & \text{ for } n=n_c\,,\\
            \frac29\, \frac{(n-5)\,n_c}{n_c-n} & \text{ for } n < n_c\,,
        \end{dcases}
\end{equation}
where $H_I/\Hrh = \left(\Tmax/\Trh\right)^{n_c/3}$ has been utilized.

We now turn to the special case $x = -3/2$ (equivalently to $k = 3/2$). As evident from Eq.~\eqref{eq:rhoR3}, this corresponds to a constant SM temperature during reheating. Thus, $x = -3/2$ marks the transition between the cases where the SM energy density grows with time during reheating $(x < -3/2)$, and where there is no period of increasing temperature $(x > -3/2)$. In this case one can write the evolution of DM number as
\begin{equation}
    \frac{dN}{da} \simeq a_I^{-3/2} \frac{\Trh^n}{H_I\, \Lambda^{n-4}}\, a^\frac72,
\end{equation}
following Eq.~\eqref{eq:BEa}, which eventually leads to a DM number
\begin{equation}
    N(\arh) \simeq \frac29 \frac{\Trh^n}{H_I\, \Lambda^{n-4}}\, a_I^3 \left[\left(\frac{\arh}{a_I}\right)^\frac92 - 1\right],
\end{equation}
at the end of reheating, again assuming $\mdm < \Trh$. Following the same prescription as before, the DM yield can be obtained as
\begin{equation}
    Y(\arh)\simeq \frac{5}{\pi^2\,\gss}\,\frac{\Trh^{n-3}}{\Hrh\,\Lambda^{n-4}}\,.
\end{equation}
Finally, the boost factor in this case reads
\begin{equation}
    B \simeq \frac29\, (n-5)\,,
\end{equation}
and is always $\mathcal{O}(1)$. This can be understood by noticing that, even if DM is constantly produced during reheating and at the same rate (the SM temperature being a constant), the DM yield suffers from the entropy dilution from the decay of the inflaton. Therefore, the production is largely dominated by late times, close to the end of reheating, and hence a small boost is expected.

\begin{figure}[t!]
    \def\sepf{0.55}
	\centering
    \includegraphics[scale=\sepf]{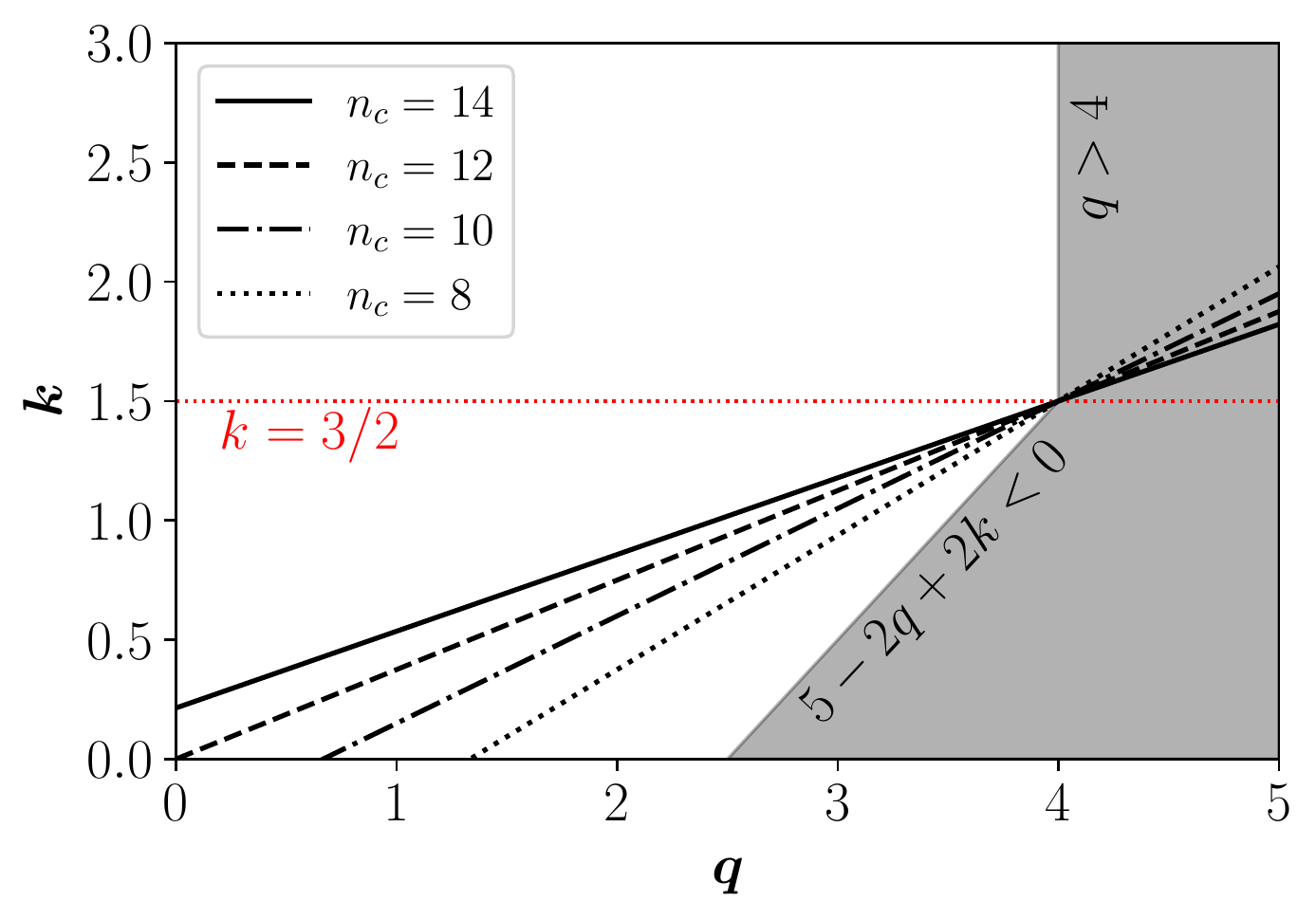}
    \caption{In the $[k,\,q]$ plane the black straight contours (cf. Eq.~\eqref{eq:nc}) correspond to different choices of $n_c:\{8\,,10\,,12\,,14\}$ from bottom right to top left. The shaded regions are disallowed from the requirement of having $q<4$ and $5-2q+2k>0$. Below the red dotted straight horizontal line at $k=3/2$ temperature of the Universe decreases during reheating, while above, the temperature increases throughout reheating (see text).
    }
	\label{fig:nc}
\end{figure}

In Fig.~\ref{fig:nc} we depict the parameter space of our interest in the $[k,\, q]$ plane, where the black-slanted straight contours correspond to $n_c=\{8,\, 10,\, 12,\, 14\}$ from bottom to top. The shaded gray region is excluded from the requirement of $q-4<0$ and $5-2q+2k>0$, which corresponds to a vanishing radiation energy density just after inflation.
Let us remind that the parameter space below (above) the line $n_c = 12$, or equivalently $x = 0$, is associated with an inflaton decay width that increases (decreases) with time.
Since, as mentioned before, a power-law enhanced boost in the DM yield is achievable only in the case where $n_c<n$, hence below these contours a sizable boost factor is viable for operators with a given $n>n_c$. 
Finally, we note that, as expected, a large boost can only appear in the case where $k < 3/2$, where the SM bath could reach a temperature $\Tmax$ much larger than $\Trh$. 

Before closing this section, we focus on the case where $\Tmax \gg \mdm \gg \Trh$, which can happen only for $x > -3/2$ (or equivalently $k < 3/2$). In this scenario, DM is produced during, but not after reheating. The DM number could be estimated by integrating Eq.~\eqref{eq:BEa} from $a_I$ to $\adm$, where $\adm \equiv a(T=\mdm)$ is given by
\begin{equation}
    \adm = \arh \left(\frac{\Trh}{\mdm}\right)^\frac{2n_c}{9}.
\end{equation}
Thus, one gets
\begin{equation}
    N(\adm) \simeq
    \begin{dcases}
        \frac29 \frac{n_c}{n_c-n}\frac{a_I^3\, \Tmax^n}{H_I\, \Lambda^{n-4}} \left[\left(\frac{\Tmax}{\mdm}\right)^{n_c-n} - 1\right] & \text{ for } n \ne n_c\,,\\
        \frac23 \frac{\arh^3\, \Trh^{n_c}}{\Hrh\, \Lambda^{n_c-4}}\, \ln\left[\left(\frac{\Trh}{\mdm}\right)^\frac{n_c}{3} \frac{H_I}{\Hrh}\right] & \text{ for } n = n_c\,.
    \end{dcases}
\end{equation}
Hence the DM yield $Y(\adm) = N(\adm)/(\adm^3\, s(\mdm))$ reads
\begin{equation}
    Y(\adm) \simeq
    \begin{dcases}
        \frac{5}{\pi^2\, \gss}\, \frac{n_c}{n_c-n}\, \frac{\mdm^{n-3}}{H_I\, \Lambda^{n-4}} \left(\frac{\Tmax}{\mdm}\right)^\frac{n_c}{3} & \text{ for } n < n_c\,,\\
        \frac{5}{\pi^2\, \gss}\, \frac{n_c}{n-n_c} \frac{\mdm^{n-3}}{\Hrh\, \Lambda^{n-4}} \left(\frac{\Tmax}{\mdm}\right)^{n - n_c} \left(\frac{\Trh}{\mdm}\right)^\frac{n_c}{3} & \text{ for } n > n_c\,.
    \end{dcases}
\end{equation}
It is important to note here that the DM yield is conserved after reheating ($a > \arh$), but not during reheating (and in particular within the range $\adm < a < \arh$), due to the entropy dilution from the inflaton decay. Then, the DM yield at the end of reheating is
\begin{equation}
    Y(\arh) = Y(\adm)\, \frac{S(\adm)}{S(\arh)} = Y(\adm)\, \left(\frac{\mdm}{\Trh}\right)^3 \left(\frac{\Trh}{\mdm}\right)^\frac{2n_c}{3}.
\end{equation}
We emphasize that in this case the DM is only produced during (and not after) reheating, and therefore the boost factor can not be defined.

\section{Scenarios with Time-dependent Inflaton Decay} \label{sec:examples}
Until now, we have discussed the consequences of time-dependent inflaton decay, especially focusing on how it can modify the DM abundance completely model-agnostic, using the generic parameterization for the inflaton decay width (cf. Eq.~\eqref{eq:inf-decay}), together with a generic DM production rate in Eq.~\eqref{eq:gamma-SM}.
In this section, we would motivate how such DM production rates can originate from a physically realizable framework. We focus on the following three situations, where DM can be produced via $i)$ 2-to-2 scattering of the bath particles through an $s$-channel graviton exchange, $ii)$ 2-to-2 scattering of the inflatons during reheating via a graviton exchange, and $iii)$ decay of the inflaton field into DM final states. The first two cases belong to pure gravitational production, where the DM particle production occurs via gravitational interaction which is described in terms of coupling of the energy-momentum tensor to the metric perturbation, where the latter is identified as the quantum field for the spin-2 massless graviton~\cite{Donoghue:1994dn, Choi:1994ax, Holstein:2006bh}. This subsequently gives rise to interactions of all matter fields with the graviton and also pair production of DM particles, satisfying the whole observed DM abundance~\cite{Garny:2015sjg, Tang:2016vch, Garny:2017kha, Tang:2017hvq, Mambrini:2021zpp, Bernal:2021kaj, Barman:2021ugy}. On the other hand, DM can also be produced directly from inflaton decay with a tiny branching fraction, if the inflaton couples to DM fields.%
\footnote{Even if the DM does not have a direct coupling to the inflaton, one cannot ignore the radiative decay of inflaton into DM~\cite{Kaneta:2019zgw}.} In the following subsections, we shall see how a time-dependent inflaton decay width influences DM production within the frameworks mentioned above.

\subsection{Gravitational UV Freeze-in}
\begin{figure}[t!]
    \def\sepf{0.53}
	\centering
    \includegraphics[scale=\sepf]{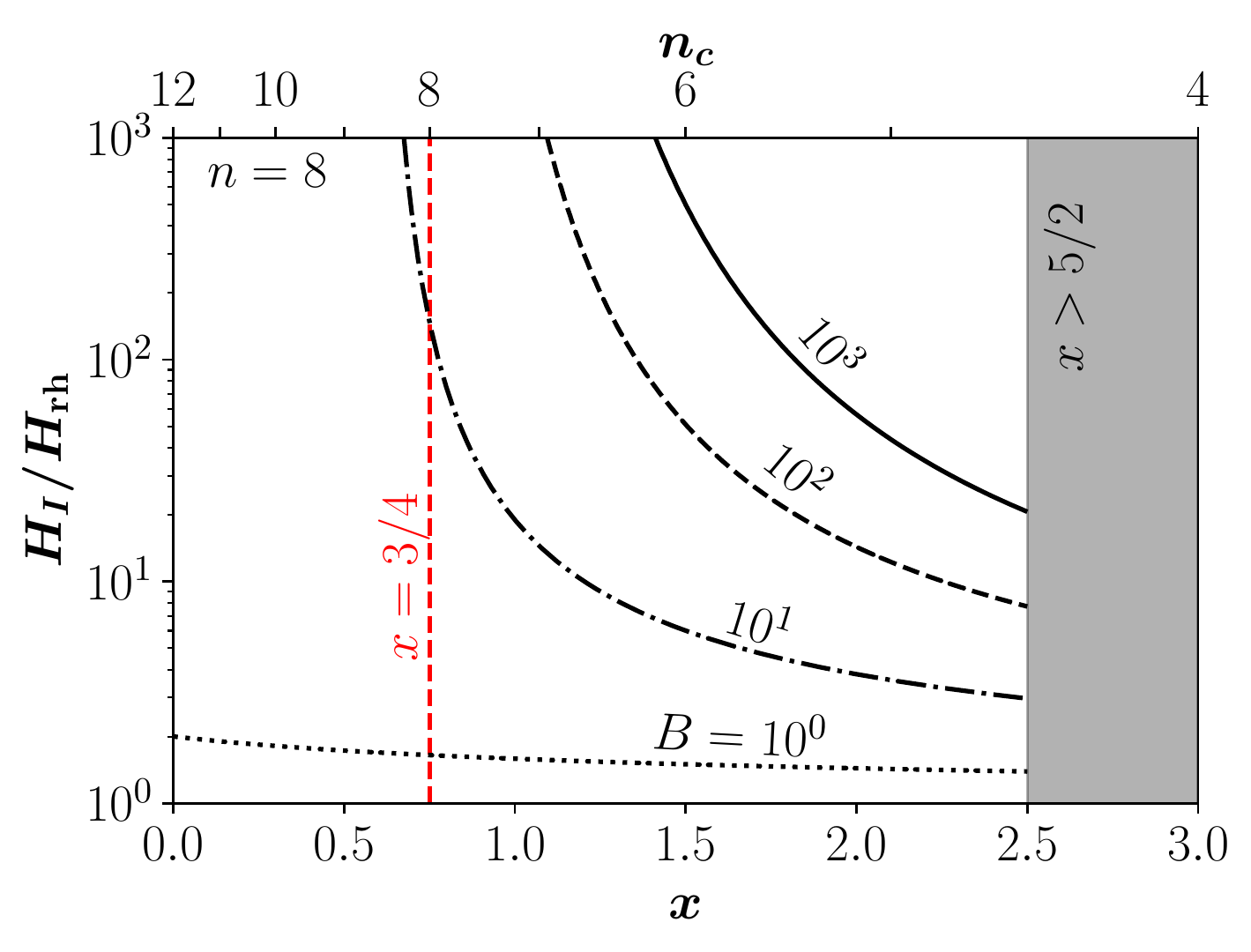}
    \includegraphics[scale=\sepf]{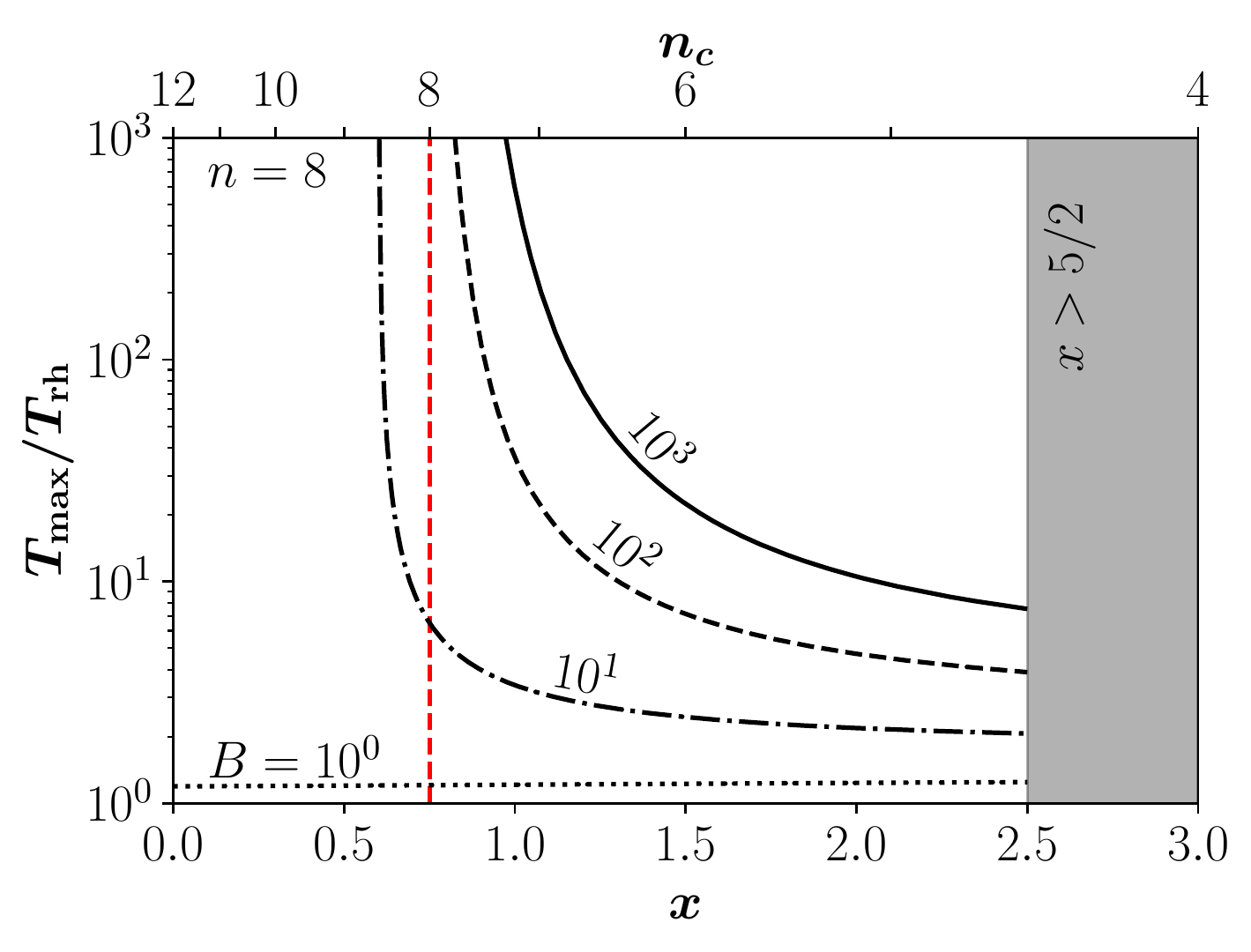}
    \caption{Contours of boost factor $B$ for the gravitational UV freeze-in scenario with SM scattering (i.e., $n = 8$), where $B=10^0$, $10^1$, $10^2$ and $10^3$ from bottom left to top right.  Power-law boosts happen for $3/4 < x < 5/2$. The gray-shaded area corresponding to $x> 5/2$ cannot be explored in the case of a negligible DM population at the end of inflation. The left and right panel correspond to the ratios $H_I/\Hrh$ and $\Tmax/\Trh$, respectively.
    } 
	\label{fig:n=8}
\end{figure}

We examine the first case where the DM is gravitationally produced via the UV freeze-in mechanism of 2-to-2 annihilation of SM particles mediated by the $s$-channel exchange of gravitons. The interaction rate density for such a process reads~\cite{Garny:2015sjg, Tang:2017hvq, Garny:2017kha, Bernal:2018qlk, Bernal:2020ili, Bernal:2021akf, Barman:2021ugy}
\begin{equation}
    \gamma = \alpha\, \frac{T^8}{M_P^4}\,,
\end{equation}
with $\alpha \simeq 1.9\times 10^{-4}$ (real scalar DM),  $\alpha \simeq 1.1 \times 10^{-3}$ (Dirac DM) or $\alpha \simeq 2.3 \times 10^{-3}$ (vector DM). Compared to Eq.~\eqref{eq:gamma-SM}, we find that this situation corresponds to $n = 8$.%
\footnote{UV freeze-in via SM scattering through the dilaton portal can give rise to $n=12$ for fermionic DM~\cite{Ahmed:2021try}.} Thus, comparing with the general form of the boost factor in Eq.~\eqref{eq:boost}, we see that the DM yield is boosted if $n_c < n = 8$, which corresponds to $x > 3/4$. This is reflected in Fig.~\ref{fig:n=8}, where contours of different choices of the boost factor are shown, for $n = 8$.
The left panel describes  the required ratio of Hubble parameter, namely  $H_I/\Hrh$ in order to yield a fixed boost factor $B$, while the right panel depicts the same with respect to $\Tmax/\Trh$. Note that these two ratios are connected by $H_I/\Hrh = \left(\Tmax/\Trh\right)^{n_c/3}$, cf. Eq.~\eqref{eq:Tmax}.
Small $\mathcal{O}(1)$ boost appear when $x < 3/4$, whereas larger boosts can happen in the opposite case.
This figure was produced using the analytical approximation in Eq.~\eqref{eq:boost}; we have checked the good agreement with the fully numerical result.

\subsection{Gravitational Inflaton Scattering}
Similarly to SM scattering, during reheating, the entire observed DM abundance can be generated by 2-to-2 annihilations of inflatons, mediated by the $s$-channel exchange of gravitons. This corresponds to the second case in our list of examples. The interaction rate density for DM production out of non-relativistic inflatons then reads~\cite{Mambrini:2021zpp, Barman:2021ugy, Bernal:2021kaj, Haque:2021mab, Clery:2021bwz, Haque:2022kez}
\begin{equation}
    \gamma = \frac{\rp^2}{512\pi\, M_P^4}\, f\left(\frac{\mdm}{m_\phi}\right),
\end{equation}
where $m_\phi$ is the inflaton mass and
\begin{equation}
    f(y) \equiv
        \begin{dcases}
            \left(y^2+2\right)^{2}\sqrt{1-y^2} & \text{for real scalars},\\
            y^2\left(1-y^2\right)^{3/2} & \text{for Dirac fermions},\\
            \frac18\sqrt{1-y^2}\left(4+4y^2+19y^4\right) &\text{for vector bosons.}
        \end{dcases}
\end{equation}
By writing Eq.~\eqref{eq:rho-inf} as a function $n_c$
\begin{equation}
    \rp(T) = 3\, M_P^2\, \Hrh^2 \left(\frac{T}{\Trh}\right)^{\frac23 n_c},
\end{equation}
one can express the reaction density as
\begin{equation} \label{eq:gammaT}
    \gamma = \frac{9}{512\,\pi}\, \Hrh^4 \left(\frac{T}{\Trh}\right)^{\frac43 n_c} f\left(\frac{\mdm}{m_\phi}\right)
     = \frac{9}{512\,\pi}\, \Hrh^4 \left(\frac{\arh}{a}\right)^6 f\left(\frac{\mdm}{m_\phi}\right).
\end{equation}
Again, comparing with Eq.~\eqref{eq:gamma-SM}, we see that in this case, $\Lambda^\frac{4\, (3 - n_c)}{3} = \frac{9}{512 \pi} \left(\Hrh\, \Trh^{-\frac{n_c}{3}}\right)^4 f$ and $n = \frac43 n_c$.
Following the same methodology as before, we can determine the DM yield at the end of reheating by solving the BEQ (namely Eq.~\eqref{eq:BEa}), that reads 
\begin{equation}
    Y(\arh) = \frac{135}{512\, \pi^3\, \gss}\, \frac{H_I\, \Hrh^2}{\Trh^3}\, f\left(\frac{\mdm}{m_\phi}\right),
\end{equation}
which is, interestingly, independent of $x$. 
This is expected from Eq.~\eqref{eq:gammaT}, where we see that the reaction density does not depend on $n_c$ expressed in terms of the scale factor. Physically, this can be realized from the fact that DM production from inflaton scattering depends on the inflaton energy density and not on the thermal distribution of the bath particles. 
Finally, we note that, as this channel is only open during reheating, a boost factor cannot be defined.

\subsection{DM from Direct Inflaton Decays}
We finally take up the case where the inflaton decay gives rise to the observed DM abundance through a small branching fraction $\Br$. In this case, the interaction rate density is
\begin{equation}
    \gamma = 2\,\Br\, \Gp\, n_\phi
    \simeq 3\, (5 - 2x)\, \Br \, \frac{M_P^2\, \Hrh^3}{m_\phi} \left(\frac{T}{\Trh}\right)^\frac{8\, (3+x)}{3+2x},
\end{equation}
where $\Br \ll 1$ is required to match the observed DM density, and therefore the formalism developed in Sec.~\ref{sec:non-inst-decay} (i.e., ignoring possible direct decays of the inflaton into DM) is not modified.
With this we obtain the DM yield at the end of reheating as
\begin{equation} \label{eq:decay_yah}
    Y(\arh) \simeq
    \begin{dcases}
        \frac32\, \frac{5-2x}{2x-3}\, \frac{\gs}{\gss}\, \Br\, \frac{\Trh}{m_\phi} \left[\left(\frac{H_I}{\Hrh}\right)^{\frac{2x-3}{3}} - 1\right] &\text{ for } x \ne \frac32\,,\\
       \frac{\gs}{\gss}\, \text{Br}\, \frac{\Trh}{m_\phi} \ln\left(\frac{H_I}{\Hrh}\right) &\text{ for } x = \frac32\,.
    \end{dcases}
\end{equation}

\begin{figure}[tb!]
    \def\sepf{0.53}
	\centering
    \includegraphics[scale=\sepf]{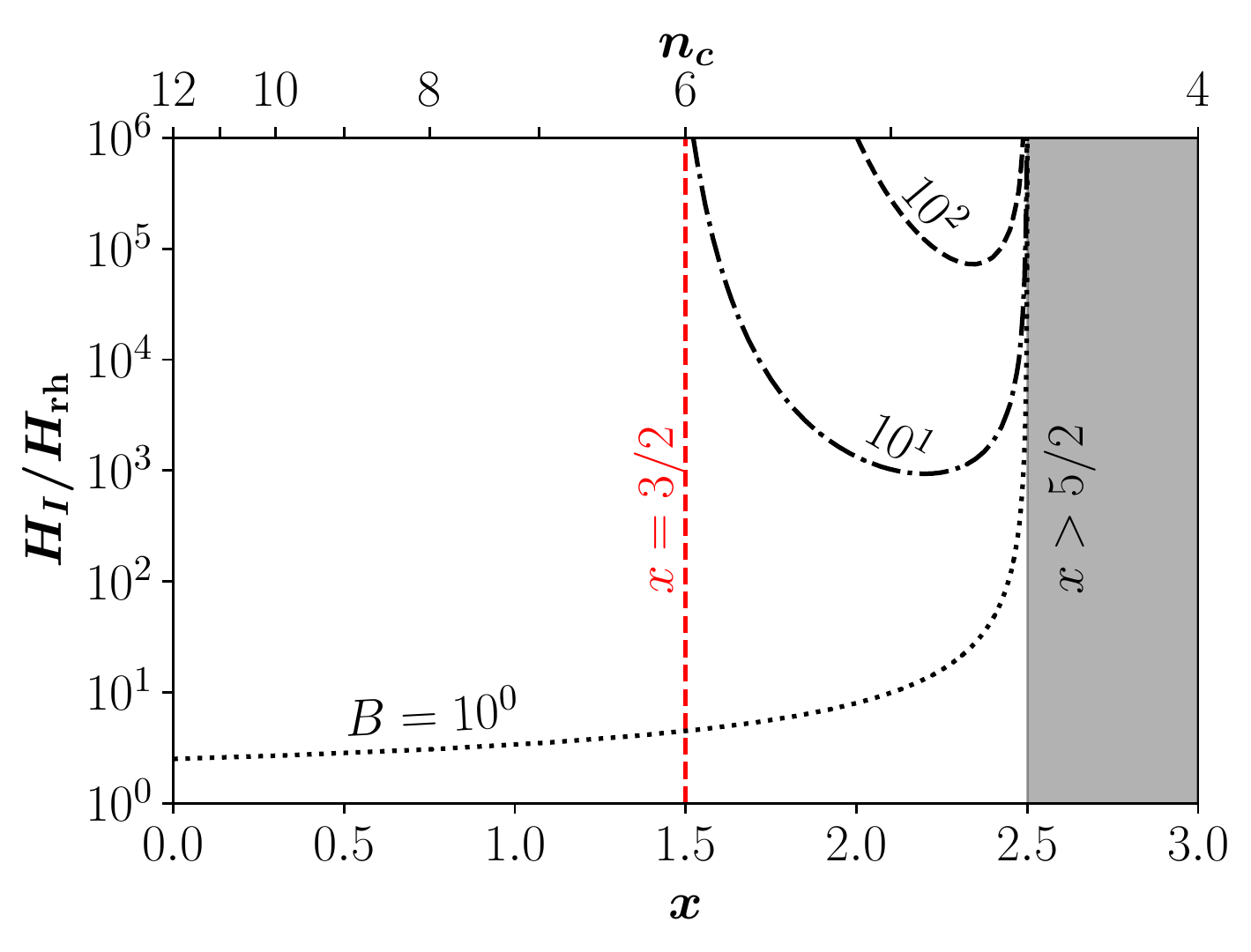}
    \includegraphics[scale=\sepf]{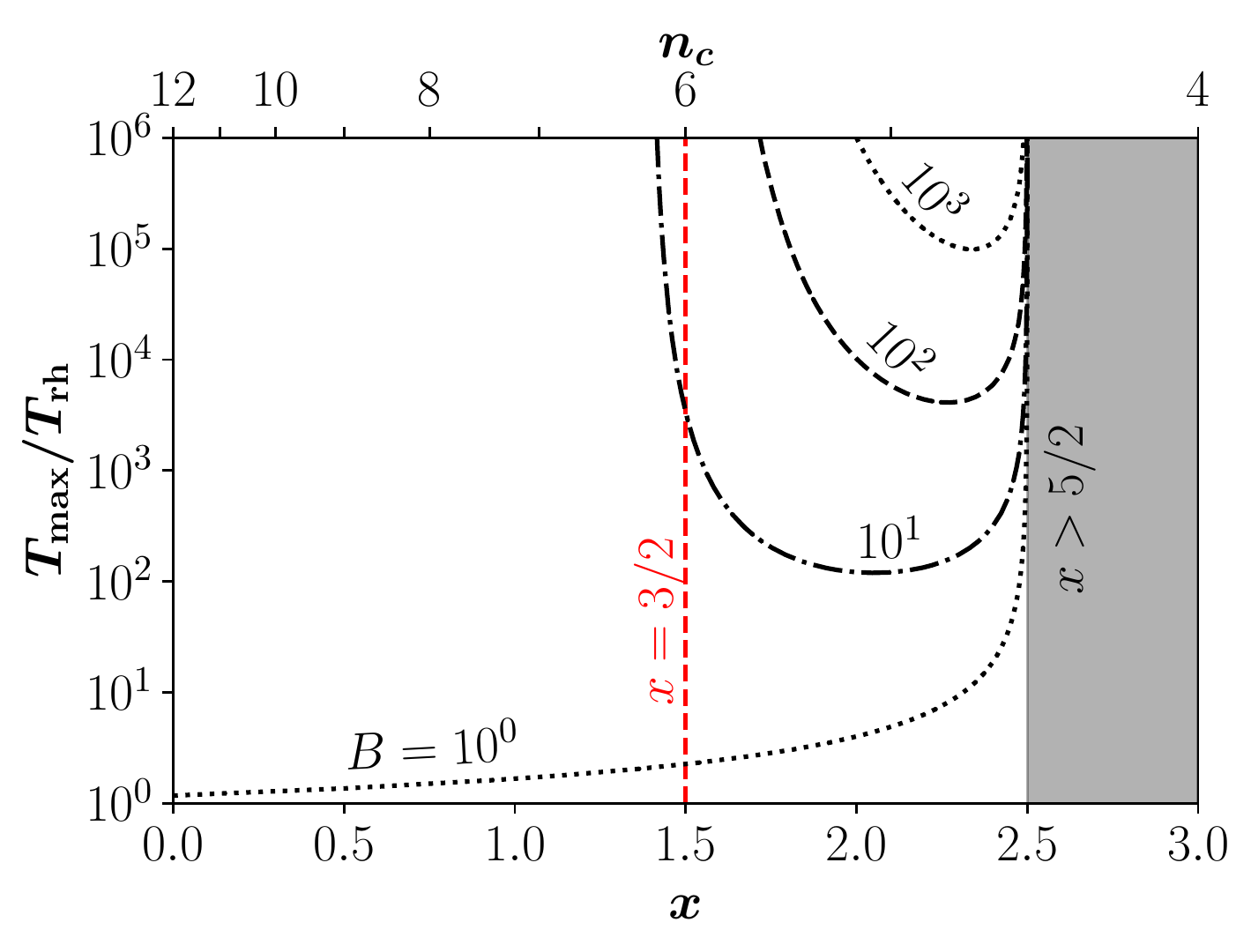}
    \caption{Contours for the boost factor for the DM production from direct decay of the inflaton, where $B=10^0$, $10^1$, $10^2$, and $10^3$ from bottom left to top right.
    The gray-shaded area corresponding to $x> 5/2$ cannot be explored in the case of a negligible DM population at the end of inflation. 
    The left and right panel correspond to the ratios $H_I/\Hrh$ and $\Tmax/\Trh$, respectively.}
	\label{fig:DecInf}
\end{figure}

Taking into account that $Y_0^\text{decay} = \frac32\, \frac{\gs}{\gss}\, \Br\, \frac{\Trh}{m_\phi}$ is the DM yield in the instantaneous reheating approximation (see, e.g. Ref.~\cite{Bernal:2021qrl}), the boost in the DM yield reads
\begin{equation}\label{eq:B-decay}
    B \equiv \frac{Y(\arh)}{Y^{\text{decay}}_0} \simeq
    \begin{dcases}
        \frac{5-2x}{3-2x} &\text{ for } x < \frac32\,,\\
        \frac{2}{3}\ln\left(\frac{H_I}{\Hrh}\right) &\text{ for } x = \frac32 \,,\\
        \frac{5-2x}{2x-3} \left(\frac{H_I}{\Hrh}\right)^\frac{2x-3}{3} &\text{ for } x > \frac32\,.
    \end{dcases}
\end{equation}
Once again, for $x>-3/2$ (or equivalently $k<3/2$), one can write the boost factor in terms of the temperature ratio $\Tmax/\Trh$ as
\begin{equation}\label{eq:B-decay2}
    B  \simeq
    \begin{dcases}
        \frac{5-2x}{3-2x} &\text{ for } x < \frac32\,,\\
        \frac{8}{3+2x}\ln\left(\frac{\Tmax}{\Trh}\right) &\text{ for } x = \frac32 \,,\\
        \frac{5-2x}{2x-3} \left(\frac{\Tmax}{\Trh}\right)^\frac{4(2x-3)}{3+2x} &\text{ for } x > \frac32\,.
    \end{dcases}
\end{equation}
We note that the boost factor can be sizable, i.e., a power law in $H_I/\Hrh$ or equivalently $\Tmax/\Trh$, if $x > 3/2$. On top of that, we also need to have $x<5/2$ for an inflaton dominated epoch during reheating, as explained before.
Hence, to have a significantly large boost, a large $H_I/\Hrh$ or equivalently $\Tmax/\Trh$ is required for $3/2 < x < 5/2$, as shown respectively in the left and right panels of Fig.~\ref{fig:DecInf}.
This figure was produced using the analytical approximation in Eq.~\eqref{eq:boost}, in good agreement with the complete numerical result.

\section{Conclusions}
\label{sec:concl}
In conventional reheating scenarios, a perturbative decay of the inflaton is typically assumed, which ends at the so-called reheating temperature, whence the inflaton and the SM radiation have comparable energy densities. In standard lore, the temperature $T$ of the thermal bath decreases during reheating as $T(a)\propto a^{-3/8}$ (where $a$ corresponds to the scale factor). However, this may not have been the case for several instances in which the temperature of the Universe can either increase or remain steady during the reheating epoch. For example, it is shown that when the inflaton oscillates around a potential steeper than quadratic or  decays through higher-order operators, its decay width can feature a time dependence~\cite{Bodeker:2006ij, Mukaida:2012qn, Co:2020xaf, Garcia:2020wiy, Ahmed:2021fvt, Banerjee:2022fiw}, leading to a non-standard evolution of the SM temperature. A general parametrization of the inflaton dissipation rate can indeed capture the nontrivial period of reheating. In such a case, the time-dependent decay of the inflaton not only influences the temperature evolution of the Universe, but can also critically affect the dark matter (DM) production during reheating. 

By parameterizing the inflaton decay in such a manner, in this work, we have shown that the UV freeze-in yield of DM can have a power-law boost in the ratio of the highest temperature and the reheating temperature, namely $\Tmax/\Trh$, which is particularly important if the temperature of the thermal bath drops fast enough during reheating.
We find the evolution of DM yield is controlled by only three free parameters: $x$, that determines the decay rate of the inflaton, along with the two Hubble scales $H_I$ and $\Hrh$ (or equivalently two temperature scales $\Tmax$ and $\Trh$).
Our main results are derived from a completely model-agnostic perspective, however we also provide a few instances where the temperature-dependent inflaton decay can play a nontrivial role in determining the final DM abundance in context with physically realizable scenarios. We examine the example of DM production from 2-to-2 scattering of the inflaton and SM bath particles, where in the former case the yield remains independent of the inflaton decay dynamics. For DM production through inflaton decay, we find that a boosted DM yield requires a significantly large $H_I/\Hrh$ or equivalently $\Tmax/\Trh$ ratio compared to the SM scattering case.

\section*{Acknowledgments}
BB and NB received funding from the Patrimonio Autónomo - Fondo Nacional de Financiamiento para la Ciencia, la Tecnología y la Innovación Francisco José de Caldas (MinCiencias - Colombia) grant 80740-465-2020.
NB and OZ received funding from MinCiencias through the Grant 80740-492-2021.
NB is also funded by the Spanish FEDER/MCIU-AEI under grant FPA2017-84543-P.
The work of OZ has been partially supported by Sostenibilidad-UdeA, the UdeA/CODI Grants 2017-16286 and 2020-33177.
This project has received funding and support from the European Union's Horizon 2020 research and innovation programme under the Marie Sk{\l}odowska-Curie grant agreement No.~860881 (H2020-MSCA-ITN-2019 HIDDeN).

\bibliographystyle{JHEP}
\bibliography{biblio}

\end{document}